\documentclass[11pt,twoside]{article}
\usepackage{asp2010}

\newcommand{\be}{\begin{equation}}

\newcommand{\ee}{\end{equation}}

\newcommand{\Rn}{{\sf {R_v}}}

\def\al{Alfv\'en\ }

\def\apjl{Astrophys. J. Lett.}                
\def\aap{Astron. \& Astrophys.}                

\begin{document}
\resetcounters
\markboth{Lin, Ng, and Bhattacharjee}{Simulations of the Corona Using GPUs}

\title{Large-Scale High-Lundquist Number Reduced MHD Simulations of the Solar 
Corona Using GPU Accelerated Machines}
\author{L. Lin$^1$, C.S. Ng$^2$, and A. Bhattacharjee$^1$}
\affil{$^1$Space Science Center, University of New Hampshire, Durham, NH 03824}
\affil{$^2$Geophysical Institute, University of Alaska Fairbanks, Fairbanks, AK
99775}

\begin{abstract}
We have recently carried out a computational campaign to investigate a model
of coronal heating in three-dimensions using reduced magnetohydrodynamics
(RMHD). Our code is built on a conventional scheme using the pseudo-spectral method,
and is parallelized using MPI. The current investigation
requires very long time integrations using high Lundquist numbers, where the
formation of very fine current layers challenge the resolutions achievable
even on massively parallel machines. We present here results of a port to
Nvidia CUDA (Compute Unified Device Architecture) for hardware acceleration
using graphics processing units (GPUs). In addition to a brief discussion of
our general strategy, we will report code performance on several machines
which span a variety of hardware configurations and capabilities. These
include a desktop workstation with commodity hardware, a dedicated research
workstation equipped with four Nvidia C2050 GPUs, as well as several
large-scale GPU accelerated distributed memory machines: Lincoln/NCSA,
Dirac/NERSC, and Keeneland/NICS.
\end{abstract}

\section{Introduction} 

The Reduced Magnetohydrodynamics Coronal Tectonics code (RMCT) solves 
the reduced MHD equations in three dimensions using a standard pseudo-spectral semi-implicit 
scheme and predictor-corrector time stepping. The code has been 
tailored for studying the role of magnetic reconnection and singular 
current layers in the heating of the solar corona. Within the framework 
of Parker's model of coronal heating \citep{Parker1972}, a recent analysis in 
two dimensions \citep{nb2008} demonstrated  that when coherence times ($\tau_{coh}$)
of imposed photospheric turbulence are much smaller than characteristic
resistive time-scales ($\tau_{R}$), the Ohmic dissipation scales independently
of resistivity. While their initial  2-D RMHD
treatment precluded non-linear effects such as instabilities and/or magnetic
reconnection they further invoked a simple analytical argument that
demonstrated that even considering these non-linear effects which would limit
the growth of $B_{\perp}$, given small enough $\tau_{coh}$, the same
insensitivity to resistivity would be recovered. The RMCT code has been used 
in a computational campaign to  extend this analysis to three dimensions 
\citep{nlb2011}. Spanning three orders of magnitude in Lundquist number, 
this campaign requires very high resolution to correctly resolve dissipative 
MHD structures, and very long time integrations to obtain adequate
statistics. The high Lundquist number limit has proved particularly
challenging even when parallelized on distributed memory machines.
In this paper we present a comprehensive reprogramming of our RMHD code 
for hardware acceleration using general purpose graphics processing units
(GPUs). 

Recent years have seen the rapid emergence of graphics processing units 
as hardware accelerators for general purpose computation and high performance
computing. Computational scientists have benefited from  GPUs in fields  as diverse as
geology, molecular biology, weather prediction, high energy nuclear 
physics (lattice QCD), quantum chemistry, finance and oil exploration.

In computational plasma astrophysics, several groups have reported progress
using GPU acceleration for magnetohydrodynamics (MHD)
\citep{Wong2009,Wang2010,Zink2011}, astrophysical gyro-kinetics \citep{Madduri2011}, and particle-in-cell
simulations \citep{Stantchev2008PIC}. 

The work we describe here is most similar to that of \cite{Stantchev2009}.
Using a G80 generation (128 stream processors single precision)
NVidia GPU, compared with a single 3.0 GHz Intel Xeon using
$1024^{2}$ perpendicular resolution, they report a up to a $14\times$ speedup for a
Hasegawa-Mima equation solver and 25-30$\times$ speedup
for a pseudo-spectral RMHD code in single precision.  We describe in this paper 
a full fledged three dimensional reduced MHD  production code and report code 
performance using the latest generation Nvidia GPUs on GPU equipped
workstations as well as several distributed memory GPU accelerated machines.

\section{Reduced Magnetohydrodynamics and the Parallel Numerical Scheme} 
The RMHD equations are a simplified version of MHD applicable to systems
  where the plasma is dominated by a strong guide field such that 
  the timescales of interest are slow compared with the
  characteristic Alfv\'{e}n timescale $\tau_{A}$. 
These restrictions also imply
  incompressibility ($\nabla\cdot V=0$) and the exclusion of magnetosonic modes 
  (leaving only the shear Alfv\'{e}n modes propagating in $\hat{e}_{z}$).  
The RMHD equations were first  derived for the
  study of tokamak plasmas by \cite{KadPog1974} and \cite{Strauss1976},
which can be written in dimensionless form as
\begin{equation}
 \frac{ \partial
  \Omega}{\partial t}+[\phi,\Omega]=\frac{\partial J}{\partial z} +[A,J]+ \nu
\nabla^2_\perp \Omega,
\label{eq_momentum}
 \end{equation}
\begin{equation}
\frac{ \partial A}{\partial t}+[\phi,A]=\frac{\partial \phi}{\partial
  z} + \eta \nabla^2_\perp A, 
\label{eq_induction}
 \end{equation}
where $A$ is the flux function 
so that the magnetic field is expressed as 
$\mathbf{B=\mathbf{\hat{e_z}}+B_\perp}=\mathbf{\hat{e_z}}+\nabla_\perp A\times\mathbf{\hat{e_z}}$;
$\phi$ is the stream function so that the fluid velocity field is expressed as 
$\mathbf{v}=\nabla_\perp \phi\times\mathbf{\hat{e_z}}$;
$\Omega= -\nabla^2_{\perp}\phi$ is the  $z$-component of the vorticity;
$J = -\nabla^2_{\perp}A$ is the  $z$-component of the current density;  
and the  bracketed terms are Poisson brackets such that, for example, $[\phi,A]\equiv\phi_yA_x-\phi_xA_y$
  with subscripts here denoting partial derivatives. 
 The normalized viscosity $\nu$ is  the inverse of the Reynolds number $\Rn$, 
and resistivity $\eta$ is  the inverse of the Lundquist number $S$.
The normalization adopted in equations (\ref{eq_momentum}) and (\ref{eq_induction}) is such that 
  the magnetic field is in the unit of  $B_z$ (assumed to be a constant in RMHD);
velocity is in the unit of $v_A = B_z/(4\pi\rho)^{1/2}$ with a constant density $\rho$; 
length is in the unit of the transverse length scale $L_\perp$; 
time $t$ is in the unit of $L_\perp / v_A$;
$\eta$  is in the unit of $4\pi v_A L_\perp /c^2$; 
and $\nu$  is in the unit of $\rho v_A L_\perp$.  



The numerical scheme we employ in this work was adapted from
\cite{ls1994} and \cite{Long1993PhD}. The simulation domain 
is a rectangular cartesian box of ($L_{z}\times L_\perp\times L_\perp$),
permeated by guide-field $B_{z}$ line-tied at both ends representing the
photosphere. Time integration is performed with a second order predictor
corrector method. Perpendicular dimensions are bi-periodic for a
pseudo-spectral scheme using  standard two-thirds rule de-aliasing and in the
parallel dimension a second order finite difference method is used. 
The original was written in Fortran90 and parallelization is accomplished by domain
decomposition in $\hat{e}_z$  using MPI. Typical resolutions used for our 
coronal heating scaling study range from $64^{2}\times32$ up to $1024^{2}\times128$ 
and as with many pseudo-spectral schemes, the 2D FFTs dominate the
computational burden typically consuming more than $80\%$ of computation
time for the resolutions we target.

\section{Numerical Results on Parker's Model of Coronal Heating}

It is crucial to the scaling study that we obtain good statistics averaging
over time evolution in statistical steady state. As with previous long time integration studies of
the Parker model, the runs are started with a vacuum potential field.
After a time of the order of the
resistive diffusion time, the system will evolve to a statistical steady
state.

The range of $\eta$ has been extended to lower values (with $\tau_c  = 10 \ll \tau_r$)
  for about an order of magnitude 
  as compared with the study in \citep{ls1994} which stopped at $\eta \sim 10^{-3}$.
This extension of course requires significant increase in resolution, 
with our highest resolution case at $512^2 \times 64$ so far, 
as compared with $48^2 \times 10$ in \citep{ls1994}.
The main difficulty in performing these simulations is the requirement to run up to hundreds 
  or even thousands of  \al times in order to obtain good statistics of the average quantities 
  under  the driving of random boundary flow. 
  
\begin{figure*}[tc]
\centering
\includegraphics[scale=0.46]{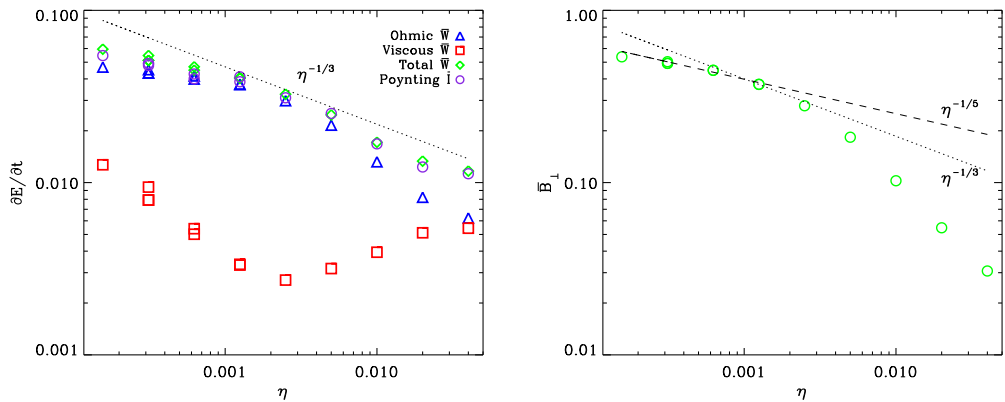}
\caption{(a) Average energy dissipation rate for different values of $\eta$. $\triangle$ is Ohmic dissipation, 
  $\Box$ is viscous dissipation, 
  $\Diamond$ is the total of the two, 
  and $\bigcirc$ is the footpoint Poynting flux. 
(b) Average perpendicular magnetic field strength for different values of $\eta$.}
\label{3dplots}
\end{figure*} 
  
 Fig.~\ref{3dplots} shows some of the scaling results we obtained so far. 
In Fig.~\ref{3dplots} (a), the time-averaged Ohmic dissipation rate $\bar{W}_\eta$ 
  (at the saturated level) for different $\eta$ is plotted in triangles,
while the viscous dissipation rate $\bar{W}_\nu$ are plotted in squares.
Note that $\bar{W}_\nu \ll \bar{W}_\eta$ in general.
The time-averaged Poynting flux $\bar{I}$ is also plotted in the same graph in circles.
It is supposed to be of the same value as $\bar{W}$ theoretically, 
and we do see that the differences between these two quantities 
  are generally small in our numerical results,
  indicating acceptable accuracy.

Due to the fact that we are doing 3D simulations,
and that we need to simulate for a long time to obtain good statistics,
so far we have only been able to extend the value of $\eta$ to about an
  order of magnitude lower, as compared with similar studies in  \citep{ls1994}.
Nevertheless, we can see already that below $\eta \sim 10^{-3}$, 
  there is a significant deviation 
  from the scalings obtained in \citep{ls1994}, who showed by numerical results and scaling analysis 
  that both $\bar{W}$ and $\bar{B_{\perp}}$ should scale with $\eta^{-1/3}$
  in the small $\eta$ limit.
We have added dotted lines in Fig.~\ref{3dplots} (a) and (b) showing 
  the $\eta^{-1/3}$ scaling.
For more details of these numerical results, 
  as well as an analysis showing the transition of scaling behavior from \citep{ls1994} to ours,
  please see \citep{nlb2011}. 

\section{GPU implementation}

CUDA is a parallel computing engine developed specifically for general
purpose applications using NVidia GPUs. It allows programmers to leverage 
the high-throughput power of GPUs by programming in the familiar
ANSI C language  rather than resorting to reverse engineering native graphics 
languages (such as Open GL or Cg) to perform general purpose and scientific
computations. 
The CUDA model allows programmers to delegate serial tasks required by the
CPU by usual C code while extensions to C are provided for programming GPUs
to exploit data parallelism. CUDA (now version 4.0 as of this writing)
provides libraries for basic linear algebra (CUBLAS), sparse matrices
(CUSPARSE), random number generation  (CURAND), standard templates (THRUST),
and fast Fourier transforms (CUFFT) as well as tools for profiling (Compute
Visual Profiler) and debugging (CUDA-gdb). In the present study, we follow 
\cite{Stantchev2009} (and most all the studies mentioned previously) in 
adopting CUDA for our acceleration project.

\begin{table}[t]
\centering
\begin{tabular}{|l|c|c|c|c|}
\hline
Name & CPU & Nodes & GPUs & Network\\
\hline
Carver/NERSC  & Nehalem(8 core)2.67 Ghz & 400 & None & 
QDR\\
Lincoln/NCSA & Harpertown(4 core)2.33Ghz & 192 & 96 x S2070
& SDR\\
Dirac/NERSC & Nehalem(8 core)2.67 Ghz & 44 & 44 x C2050 & QDR \\
UAF Workstation & Gulftown(12 core)2.8 Ghz & 1
& 4 x C2050 & - \\
Keeneland/NICS & Westmere(6 core)2.67 Ghz & 120 & 360 x C2070 & QDR\\
\hline
\end{tabular}
\caption{Specifications for Carver/NERSC and several GPU accelerated machines.}
\label{tabl}
\end{table}



Our strategy for a comprehensive reprogramming of RMCT for GPU acceleration
with CUDA can be summarized as follows: $\mathbf{[1]}$ Perform FFTs using the
CUFFT library. The library is about an order of magnitude faster than our
original FFT implementation when not considering CPU-GPU memory transfers 
but only several times faster when including them. We aim therefore to maximize 
the number of FFTs per memory transfer, and to perform intermediary tasks on
the GPU. $\mathbf{[2]}$ Recycle memory of intermediate quantities. There is
limited memory available on the GPU board so  one must adequately budget
memory allocated on board in such a manner as to  observe $\mathbf{[1]}$. 
$\mathbf{[3]}$ Write simple kernels for point-wise arithmetic. Point-wise arithmetic is
the bread and butter of stencil-operation based 
MHD codes, and CUDA implementations of such codes have been published (\cite{Wong2009} 
\cite{Wang2010} and \cite{Zink2011}). It is tempting to say that such operations take
a back seat in the  current pseudo-spectral application to FFTs, but as we
have seen Amdahl's law would require that these kernels also see full consideration.
$\mathbf{[4]}$ Preserve the underlying MPI decomposition. We are dealing here with 
two levels of parallelization, the first being the domain decomposition  
in $z$ and the second being the massive parallelism afforded by GPUs. The
wide availability of multi-GPU workstations and GPU accelerated distributed 
memory machines warrants the pursuit of both parallelization methods in tandem. 
For this code, we pair one CPU core to one GPU, each core responsible for 
one sub-cube in the the $z$ domain decomposition scheme.

\section{GPU Code Performance and Discussion} 

\begin{figure*}[tc]
\centering
\includegraphics[scale=0.69]{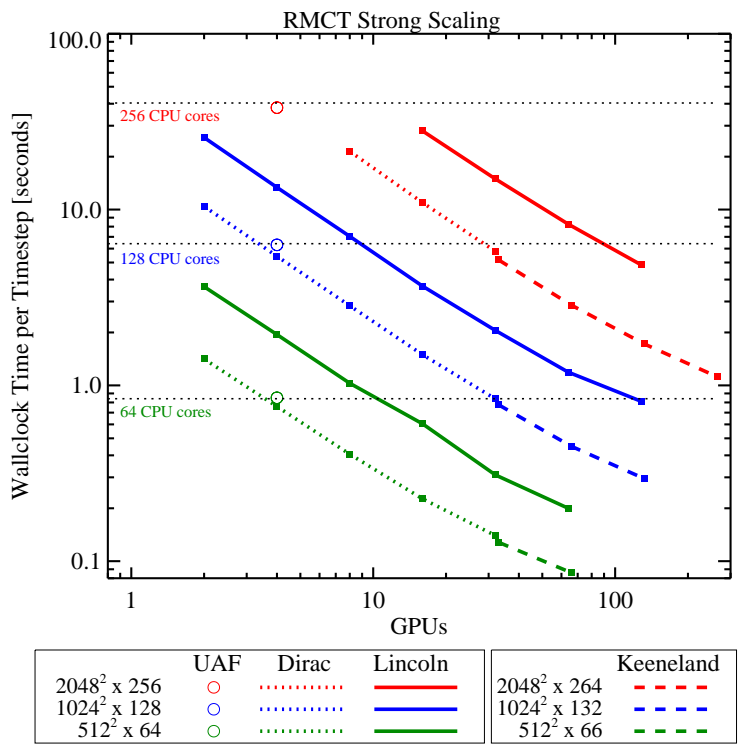}
\caption{Strong scaling of RMCT CUDA on various large scale distributed
  memory machines. \label{scal}}
\end{figure*}

Figure~\ref{scal} shows scaling results for RMCT CUDA on several 
GPU accelerated distributed memory machines. Measurements were made for each
of five machines whose relevant specifications are summarized in
Table~\ref{tabl}. For a base-line comparison we use Carver,
an IBM iDataplex mid-sized traditional CPU cluster at the National Energy
Research Scientific Computing Center (NERSC) on which we have achieved the best
performance thus far with the original version of the code. Carver
hosts a GPU testbed cluster named Dirac which accelerates each of 44 nodes with 
a single NVidia C2050 GPU. A larger dedicated GPU production cluster at the
National Center for Supercomputing Applications (NCSA) was available through
TeraGrid, and paired 96 S2070s (4 GT200 GPU module) with 192 nodes. A
dedicated GPU accelerated desktop workstation was acquired for this project
and is running at the University of Alaska Fairbanks with four C2050 GPUs. The
most powerful machine on which we have tested is the Keeneland Initial
Delivery System hosted at the National Institute for Computational Science
(NICS) and Georgia Tech. It features 120 nodes each accelerated by 3 C2070s,
and is to be expanded to a full production system for TeraGrid (now called
XSEDE) using next generation NVidia GPUs in the coming year. 

As the codes currently stand, the CUDA port on Lincoln/NCSA at full scale is
able to achieve roughly an order of magnitude speedup compared with
Carver. Considering the performance of the  code on the UAF workstation, the
chip-to-chip equivalence is roughly 4, 8, and 16. As expected, the possibility
of exposing maximal fine-grained parallelism given finer grids improves the
potential for increasing this equivalence ratio. Actually observing an
increase in this ratio then attests to the quality of the re-coding
effort. 
At full scale on Keeneland, using the Fermi class GPUs
the code is able to achieve up to a $30 \times$ speedup beyond
what was previously achieved on Carver. 

The principal caveat of the present analysis is that 
we have spent much of our time optimizing the CUDA port, and 
have used as a basis of comparison, an original Fortran/MPI 
code which remains largely the same. The results reported here therefore 
should be taken as an effective speedup for a small academic coding 
team (consisting of a professor and graduate student) going from 
one practically deployable code to another. 

\acknowledgements 
Computer time was provided by UNH (using the Zaphod Beowulf 
cluster at the Institute for the Study of Earth, Oceans and Space), as well as  a grant of HPC
resources from the Arctic Region Supercomputing Center and the DoD High Performance Computing Modernization Program.
This work is supported by NASA grants NNX08BA71G,
NNX06AC19G, a NSF grant AGS-0962477, and a DOE grant DE-FG02-07ER54832.


\bibliographystyle{asp2010}
\bibliography{thesis}

\end{document}